\documentclass[iop,english,twocolappendix,numberedappendix,appendixfloats,tighten,apj]{emulateapj}

\usepackage{amstext}
\usepackage{graphicx}
\usepackage{babel}
\usepackage{hyperref}
\usepackage{color}

\begin{document}
\title{Nature of the MHD and kinetic scale turbulence in the magnetosheath of
Saturn: Cassini observations}
\author{L. Hadid\altaffilmark{1}, F. Sahraoui\altaffilmark{1}, K. H.
Kiyani\altaffilmark{1,2}, A. Retin\`o\altaffilmark{1}, R. Modolo\altaffilmark{3}, P.
Canu\altaffilmark{1}, A. Masters\altaffilmark{4}, and M. K.
Dougherty\altaffilmark{5}}
\email{lina.hadid@lpp.polytechnique.fr}
\affil{$^1$ Laboratoire de Physique des Plasmas, CNRS-Ecole Polytechnique-UPMC,
Observatoire de Saint-Maur, France}
\affil{$^2$ Centre for Fusion, Space and Astrophysics; University of Warwick,
Coventry, CV4 7AL, United Kingdom}
\affil{$^3$ LATMOS, CNRS-UVSQ-UPMCS, Guyancourt, France}
\affil{$^4$ ISAS-JAXA, Sagamihara, Japan}
\affil{$^5$ The Blackett Laboratory, Imperial College London, SW7 2BZ, U.K.}
\begin{abstract}
       
        Low frequency turbulence in Saturn's magnetosheath is investigated using
in-situ measurements of the Cassini spacecraft. Focus is put on the magnetic
energy spectra computed in the frequency range $\sim[10^{-4}, 1]$Hz. A set
of 42 time intervals in the magnetosheath were analyzed and three main
results that contrast with known features of solar wind turbulence are
reported: 1) The magnetic energy spectra showed a $\sim f^{-1}$ scaling at
MHD scales followed by an $\sim f^{-2.6}$ scaling at the sub-ion scales
without forming the so-called inertial range; 2) The magnetic
compressibility and the cross-correlation between the parallel component of
the magnetic field and density fluctuations $ C(\delta n,\delta B_{||}) $
indicates the dominance 
of the compressible magnetosonic slow-like modes at MHD scales rather than the Alfv\'en
mode; 3) Higher order statistics revealed a monofractal (resp. multifractal)
behaviour of the turbulent flow downstream of a quasi-perpendicular (resp. quasi-parallel)
shock at the sub-ion scales. Implications of these results on theoretical modeling
of space plasma turbulence are discussed.
\end{abstract}
\maketitle
\section{Introduction} \label{sec:Intro} \emph{The solar wind} is unmatched by any
other astrophysical system in the level of details in which turbulence can be
investigated. This is due the availability of many spacecraft missions that provide
high quality field and particle in-situ measurements. The available data allowed
significant progress in understanding turbulence and energy dissipation in
collisionless magnetized plasmas. One of the most common and insightful ways of
measuring the multiscale nature of turbulence is via the Power Spectral Density
(PSD) of the turbulent fluctuations. From that point of view, it has been shown that
the magnetic energy spectrum in the solar wind is generally characterized by at
least four different dynamical ranges of scales. First is the
energy-containing range that follows a scaling $\sim f^{-1}$, which is observed
essentially in the fast solar wind and thought to be filled by uncorrelated
random-like fluctuations that may originate from reflected waves in the solar
corona~\citep{Bavassano1982,Velli1989}. The second region is the so-called inertial
range with a scaling $\sim f^{-5/3}$ or $\sim f^{-3/2}$ thought to originate from
nonlinear interactions between counter-propagating incompressible Alfv\'enic 
wave-packets  transferring the energy down to shorter
wavelength~\citep{Iroshnikov1964,Kraichnan1965}. This spectrum terminates with a
breakpoint occuring near the ion gyro-scale or inertial length scale which is
generally followed by a steeper power-law spectrum $f^{-\alpha}$ at the sub-ion
scales with a broader range of slopes, $\alpha\in[-2.3,-3.1]$, where the magnetic energy starts to dissipate into particle
heating~\citep{Goldstein1994,Leamon1998,Alexandrova2008,Hamilton2008,Sahraoui2009,Kiyani2009a}.
As the cascade approaches the electron scale, the spectrum steepens again, which is
interpreted as due to dissipation of the remaining magnetic energy into electron
heating via Landau damping of Kinetic Alfv\'en Wave (KAW) turbulence~\citep{Leamon1998b,Leamon1999,Hollweg1999,Howes2006,Sahraoui2009}. Due to instrumental limitations, the actual scaling at
sub-electron scales and the fate of the energy cascade remain open questions (see
discussions in~\cite{Sahraoui2013}).

Turbulence in the terrestrial magnetosheath is more complex than in the solar wind
as different waves and instabilities can be generated by, e.g., temperature
anisotropy generally observed behind the bow shock. Moreover, boundaries such as the
magnetopause and the shock may influence some of the turbulence properties (e.g., its spatial anisotropy)~\citep{Russell1990,Bavassano2000,Sahraoui2006,Yordanova2008}. Previous studies of
magnetic energy spectra in the terrestrial magnetosheath showed some similarities
with the solar wind: the presence of the Kolmogorov  spectral index $-5/3$ at MHD
scales~\citep{Sundkvist2007,Alexandrova2008} and a broad range of slopes, $[-2.5,
-3]$, at sub-ion scales~\citep{Czaykowska2001,Huang2014}. Some differences seem to
exist regarding the scaling at sub-electron scales~\citep{Huang2014}.

In planetary systems other than Earth, turbulence is much less explored. For
turbulence studies, there is at least one major interest in investigating planetary
magnetospheres: they offer access to a broader range of plasmas parameters that are
not available in the near-Earth space~\citep{Papen2014,Tao2015}. This is the case,
for example, of the Alfv\'en Mach number, relevant for the physics of shocks and
compressible turbulence, which can reach values as high as $\sim 100$ near Saturn~\citep{Masters2013}. The reason is that the magnetic field magnitude and the density
fluctuations decrease with different scaling laws whereas the solar wind speed stays
relatively constant~\citep{Masters2011}. Another interest is to understand the role
that the planet's satellites (e.g., Io for Jupiter) may play in  modifying locally
the turbulence properties through different plasma processes and instabilities that
the planet-moon coupling  may
generate~\citep{Chust2005,Kivelson2004,Saur2004,Bagenal2007}.
Using a list of long (several hours) crossings of the Kronian magnetosheath by the
Cassini spacecraft, we investigate the properties of turbulence at MHD and sub-ion
(kinetic) scales and compare them to the previously reported ones in the solar wind
and in the terrestrial magnetosheath. We try to answer three main questions
regarding turbulence in the magnetosheath: is the $f^{-5/3}$ Kolmogorov inertial range
ubiquitous? What is the nature of the plasma mode(s) (e.g., Alfv\'enic or
magnetosonic) that dominate the cascade at different scales; iii) Do turbulence
properties depend on the local plasma parameters (e.g., the
normal angle to the shock)?
\section{Results} \label{sec:Results} 
\subsection{Statistics of the spectral slopes} 
\begin{figure}
        \begin{centering}
                \includegraphics[width=1\columnwidth]{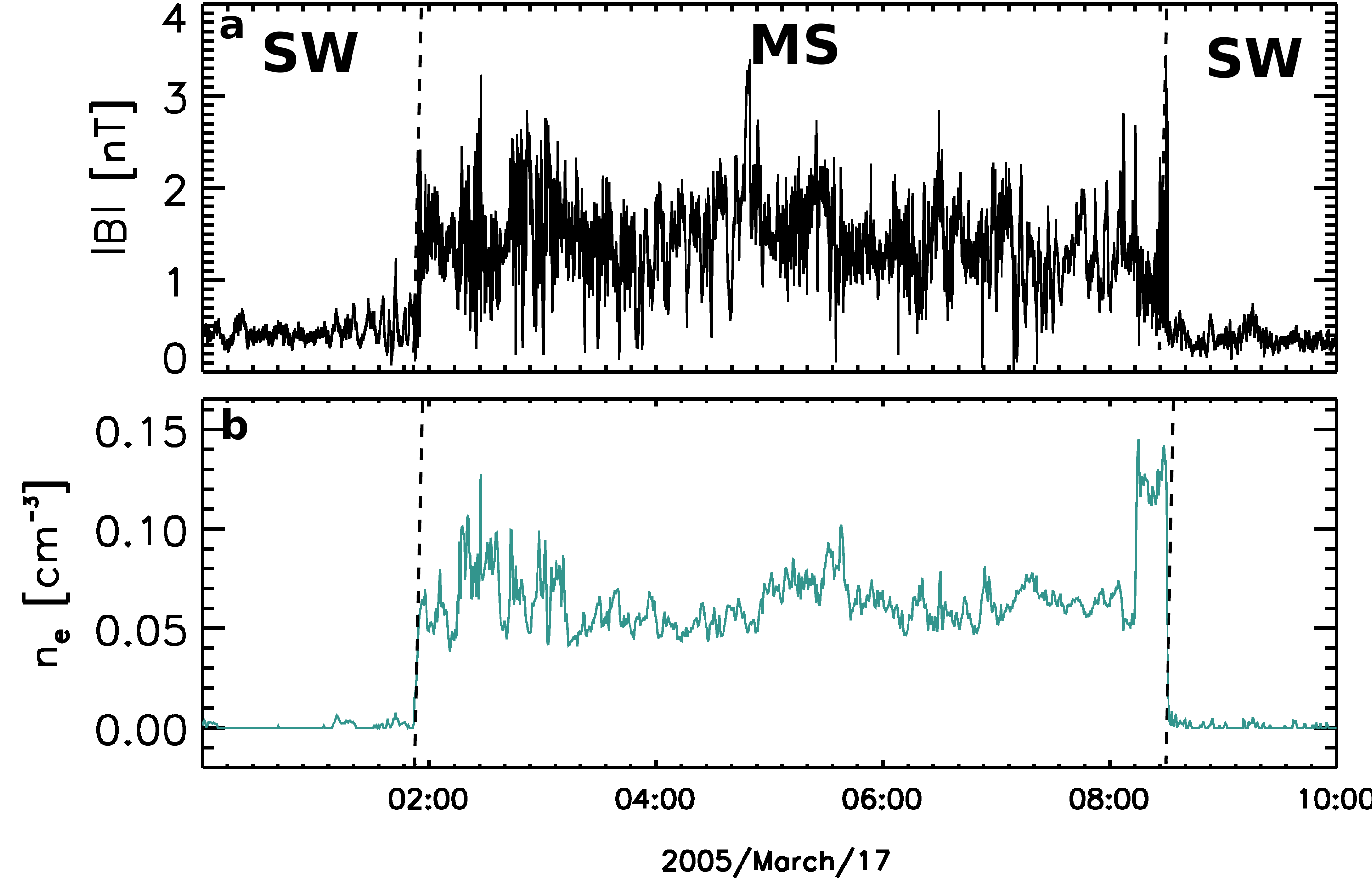} \par
        \end{centering}
        \caption{\label{fig:fig1} (a) The magnetic field modulus, (b) the electron plasma density measured by the
Cassini spacecraft in the solar wind (SW) and in the magnetosheath (MS) of
Saturn on 2005/03/17 from 00:00-10:00.}
\end{figure}

\begin{figure}
        \begin{centering}
                \includegraphics[width=1\columnwidth]{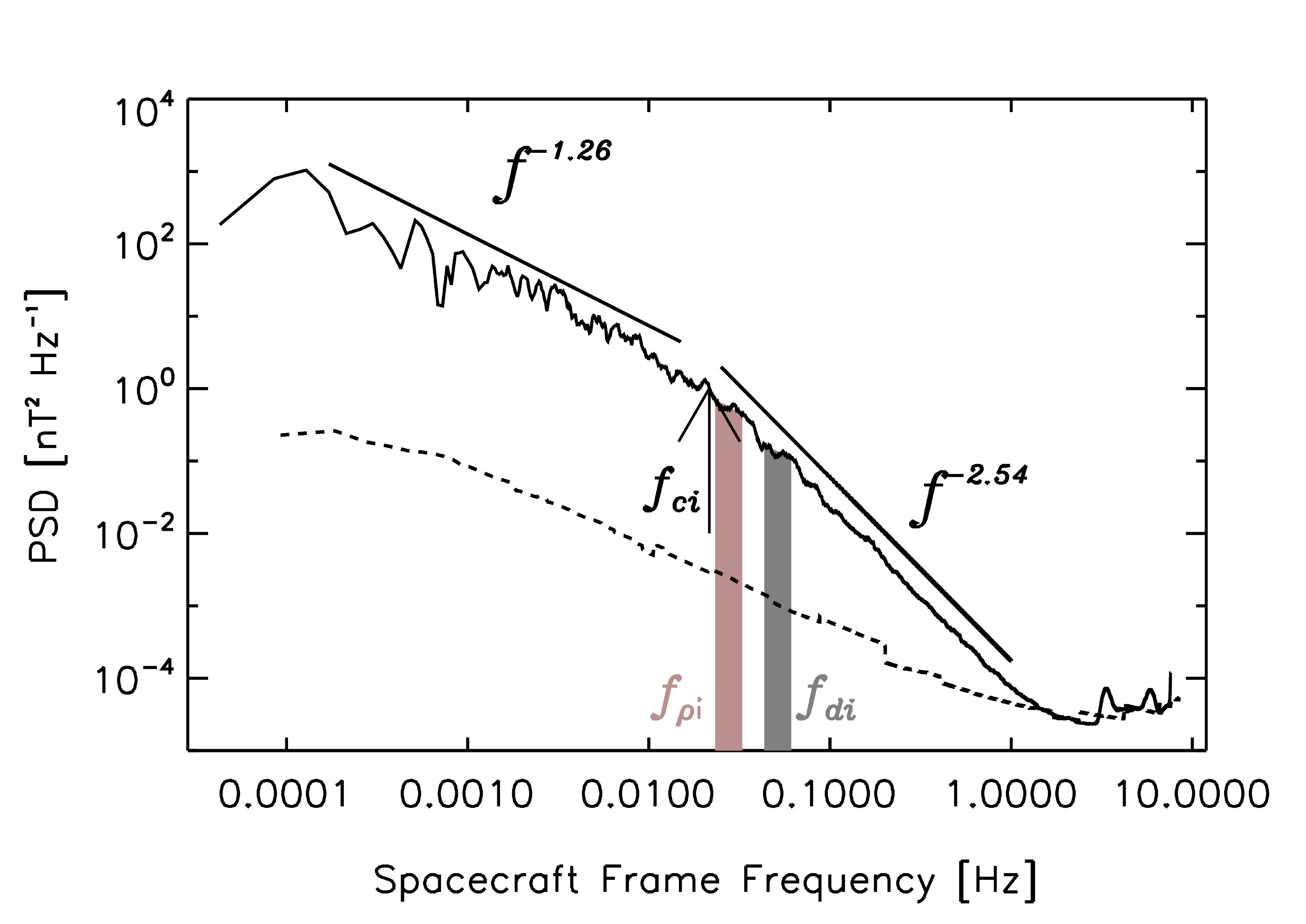} \par
        \end{centering}
        \caption{\label{fig:fig2}The power spectral density of $\delta{B}$ measured
between 02:00-08:30. The black lines are the power-law fits. The dotted
curve is a spectrum measured in the solar wind, considered here to represent
the upper bound of the sensitivity floor of FGM. The arrow corresponds to
the ion gyro-frequency, the gray and the red shaded bands indicate the Taylor
shifted ion inertial length $f_{d_i}$ and Larmor radius $f_{\rho_i}$,
respectively (the width reflects the uncertainty due to errors in estimating
the ion moments).}
\end{figure}

Figure~\ref{fig:fig1} illustrates a typical magnetosheath crossing on day March 17,
2005 at 02:00 UT and 08:30 UT (7.30 and 7.35 Local Times respectively) at a distance
of $\sim 42 R_s $ ($ 1R_s=60.268 $ km). 
From the field magnitude and the density measurements we see that the spacecraft was in the solar wind until
about 0200 UT, when it encountered the bow shock and entered into the magnetosheath where the field strength and the density increased significantly. The magnetic field data, sampled at $32Hz$, were measured by the Flux Gate
Magnetometer (FGM) sensor from the Cassini MAG experiment~\citep{Dougherty2004}. The
FGM is mounted halfway along the 11-m spacecraft boom to minimize the interference
from the spacecraft-generated electromagnetic fields. The ion and electron moments
were measured by the Cassini Plasma Spectrometer (CAPS)~\citep{Young2004}.
Cassini being a non-spinning spacecraft and the CAPS sensor having a limited Field-of -View (FoV), a careful handling of the thermal
ion population is required because the ion thermal speed is smaller than the bulk flow speed. In fact the moments are not reliable when the bulk of the plasma flow is not in the FoV of the ion instrument ~\citep{Thomsen2010,Romanelli2015}. However, since the electrons have a thermal speed
that is larger than the bulk fow speed downstream the bow
shock, the previous condition can be relaxed, and the electron can be assumed isotropic (at least on large time scales that we consider in this work) \citep{Lewis2008}. This implies that electron moments, computed from the ELS instrument,
would have less uncertainties than the ion ones. For that reason, we use the
electron density measured by ELS as the plasma density under the assumption of
quasi-neutrality $n_i\sim n_e\sim n$ (Fig.~\ref{fig:fig1}-b). 
Figure~\ref{fig:fig2} 
 the PSD of the magnetic field fluctuations computed using
a windowed Fast Fourier Transform
(FFT). The spectrum shows two ranges of scales with distinct power-laws: $\sim
f^{-1.26}$ at low frequencies ($f<10^{-2}$Hz) and $\sim f^{-2.54}$ at higher
frequencies. This observation shows a striking result: turbulence transits directly
from the ``energy containing scales" into the ion kinetic scales, without forming
the so-called Kolmogorov inertial range with a scaling $5/3$ (the terminology of
{\it energy containing scales} is borrowed from solar wind turbulence). The spectral
break is closer to the local ion gyro-frequency than to the Taylor-shifted ion
inertial length $f_{d_i}=V_{f}/2\pi d_i$ and Larmor radius $f_{\rho_i}=V_{f}/2\pi
\rho_i$ ($V_f\sim 300 $ km/s is the average flow speed, $T_{i}\sim258$ eV, $ B_0\sim1.4$ nT, $n_{e}\sim0.06$ $cm^{-3}$ and $\beta_{i}\sim 3.3 $). The reason might be that these latter are subject to
higher uncertainties due to errors in estimating the plasma parameters using the ion
moments from the CAPS instrument. 
\begin{figure}
        \begin{centering}
                \includegraphics[width=1\columnwidth]{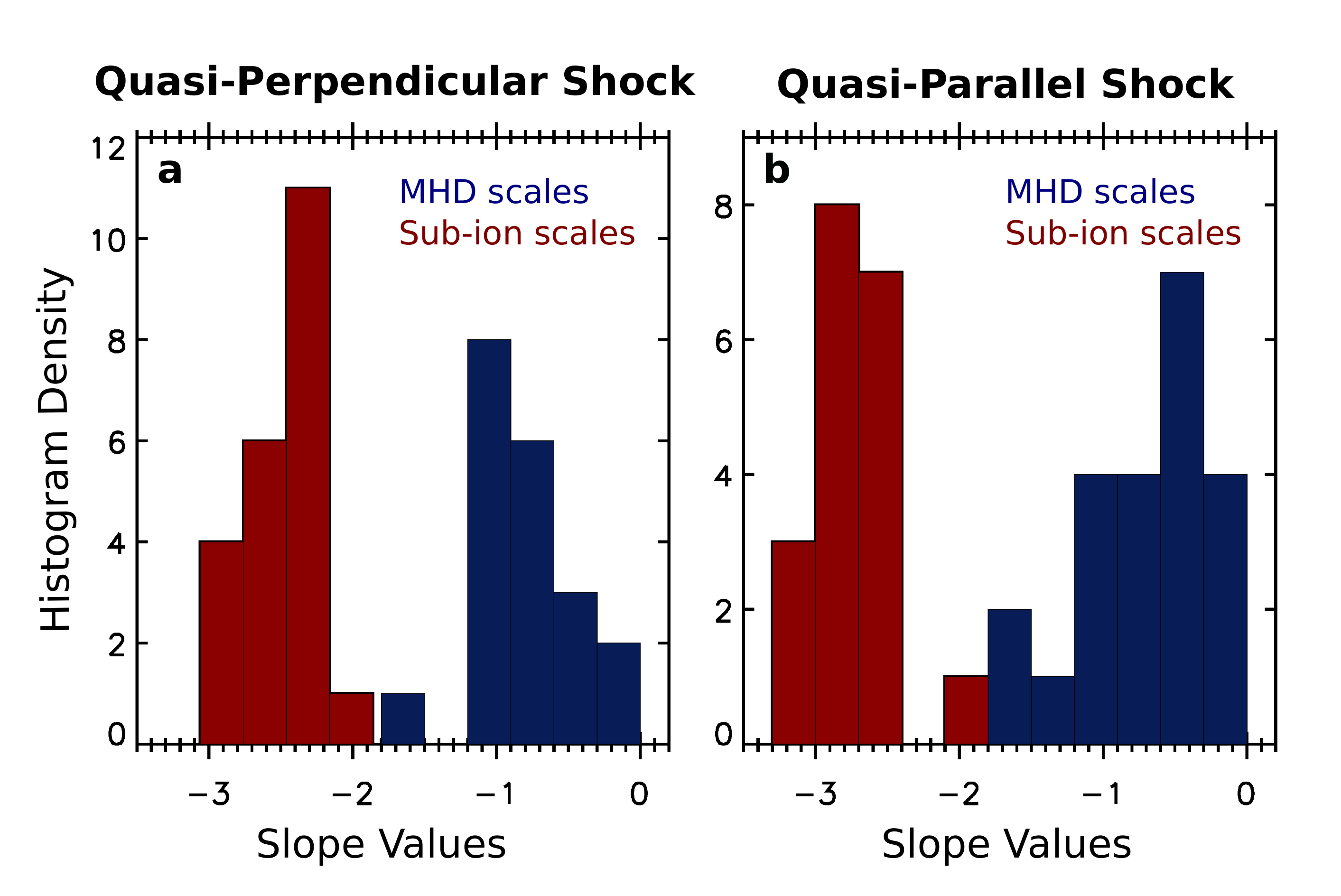} \par
        \end{centering}
        \caption{\label{fig:fig3}Histograms of the spectral slopes at MHD and sub-ion (kinetic) scales downstream of quasi-perpendicular (a) and quasi-parallel Shock (b).}
\end{figure}
To confirm the absence of the Kolmogorov $f^{-5/3}$ spectrum in the magnetosheath, 
we analyzed a list of 42 other time intervals between 2004 and 2007, for a
quasi-parallel and quasi-perpendicular shocks separately. For most of the time
intervals we identified the structure of the shock by checking the angle
$\theta_{Bn}$ between the interplanetary magnetic field and the normal to
the shock estimated using a semi-empirical model of the global shock surface~\citep{Went2011}: $\theta_{Bn} > 45^\circ$
indicates a quasi-perpendicular shock whereas $\theta_{Bn} \leq 45^\circ$ indicates a
quasi-parallel one (in few cases, quasi-perpendicular shocks are simply identified
by a sharp gradient in the magnetic field and the plasma measurements). The
results shown in Figure~\ref{fig:fig3} confirm statistically the absence of the Kolmogorov
spectrum at MHD scales: the bulk of the spectra had slopes near $-1$ in particular
for quasi-perpendicular shocks~\citep{Czaykowska2001}. The histogram of the slopes
at sub-ion scales peaks between $[-2.5, -3]$ in general agreement with previous
results reported in the solar wind and the
magnetosheath~\citep{Sahraoui2013,Huang2014}. A slight indication that steeper
spectra are observed behind quasi-parallel shocks can also be seen.
\subsection{ Nature of the turbulent fluctuations at the MHD and kinetic scales} 
\begin{figure}
        \begin{centering}
                \includegraphics[width=1\columnwidth]{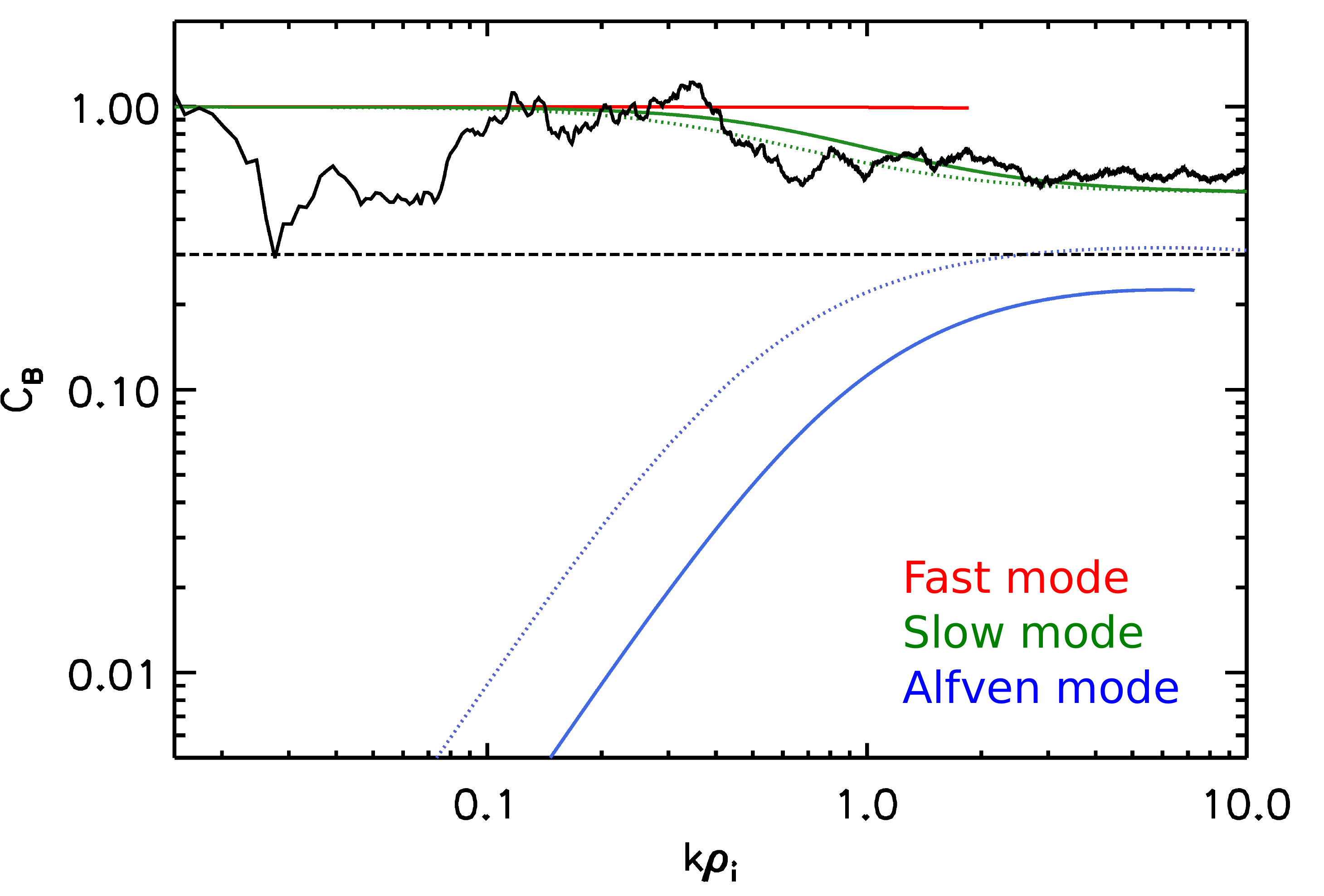}\par
        \end{centering}
        \caption{\label{fig:fig4} Comparison between theoretical magnetic compressibilities,
computed from the linear solutions of the compressible Hall-MHD (color
dotted line) and of the Vlasov-Maxwell equations (colored solid line) for
$\beta=1 $ and $\Theta_{kB}=87^\circ$, with the observed one
from the data of Fig.~\ref{fig:fig1} (02:00-08:30) (solid black curve). The
Taylor hypothesis was used to convert the frequencies in the spacecraft
frame into wavenumber. The red, green and blue curves correspond
respectively to the theoretical fast, slow and KAW modes. The horizontal
dashed black line at $C_{B}=1/3$ indicates the power isotropy level.}
\end{figure}

To identify the nature of plasma modes that carry the energy cascade from the
energy-containing to the sub-ion scales, we use the magnetic compressibility $C_B$
given by the ratio between the PSDs of the parallel magnetic field component and the
magnetic field magnitude (parallel is w.r.t. the background magnetic field $B_{0}$)
\citep{Gary2009a,Salem2012}:
\begin{equation}
C_B(f)=\frac{|\delta B_{\parallel}(f)|^{2}}{|\delta B_{\parallel}(f)|^{2}+|\delta
B_{\perp}(f)|^{2}} \label{1}
\end{equation}
Indeed, from linear theory, the Alfv\'en and the magnetosonic modes are known to
have very different profiles of the magnetic compressibility~\citep{Sahraoui2012}.
This can allow us to verify easily the dominance (or not) of the Alfv\'enic
fluctuations in our data~\citep{Podesta2012,Kiyani2013a}. We computed the
theoretical magnetic compressibilities from the linear solutions of compressible
Hall-MHD~\citep{sahraoui2003} and from the Maxwell-Vlasov equations using the WHAMP
code~\citep{Ronnmark1982}. For the sake of simplicity, we keep using the 
terminology of the MHD slow and fast modes at kinetic scales even if it may be
inadequate (because of possible crossings between different dispersion branches).
Since the slow mode is heavily damped in kinetic theory at finite
$\beta_i$~\citep{Ito2004,Howes2009}, we used the limit $\beta_i=0$ and $\beta_e=1$
(therefore, $\beta=\beta_e+\beta_i=1$) to suppress the ion Landau damping and thus
to capture the slow mode solution down to the scale $k\rho_i \geq1$. In order to
compare to spacecraft observations, the knowledge of the three components of the
${\bf k}$ vector (or equivalently, the propagation angle $\Theta_{{\bf kB}_o}$ and
the modulus $k$) from the data is required. However, unambiguous determination of
those quantities requires having multi-spacecraft data that is not available in
planetary magnetospheres other than Earth~\citep{Sahraoui2006}. Therefore, we use the
Taylor frozen-in-flow hypothesis, which assumes that the fluctuations have slow
phase speeds w.r.t the flow speed, to infer the component of the ${\bf k}$ {\it
along the flow direction}, i.e. $\omega_{sc}\sim {\bf k.V}_f \sim kV_f$. Under the assumption that turbulence is strongly
anisotropic, i.e. $k_\perp>>k_\parallel$, which is supported by previous
observations in the magnetosheath~\citep{Sahraoui2006,Mangeney2006} and in the solar
wind~\citep{Sahraoui2010}, the estimated wavenumber component along the flow is equivalent to $k_\perp$
for data intervals when $\Theta_{{\bf V}_f{\bf
B}_0}\sim 90^\circ$. In the present data we estimated
$\Theta_{{\bf V}_f{\bf B}_0}\sim 87^\circ$, which we used to compute the theoretical
solutions of Fig.~\ref{fig:fig4} assuming $\Theta_{{\bf V}_f{\bf
B}_0} \sim \Theta_{{\bf k B}_0}$. Nevertheless, we performed a parametric
study (not shown here) and verified that the magnetic compressibilities of the
compressible Hall-MHD solutions keep the same profile (but change its magnitude)
when varying  $\beta$  in the range $[0.2, 100]$ for a fixed $\Theta_{\bf {kB_o}}=
87^\circ$, and when varying $\Theta_{\bf {kB_o}}$ from quasi-parallel to
quasi-perpendicular angles for $\beta=1$. Another consequence of using the Taylor
hypothesis when $\Theta_{{\bf V}_f{\bf B}_0}\sim 90^\circ$ is that the perpendicular
component of the fluctuation $\delta B_\perp$ in Eq.~\ref{1} is reduced to the
component perpendicular to both ${\bf V}_f$ (or {\bf k}, to
fulfill ${\bf k}.{\bf \delta B}=0$) and to ${\bf B}_0$~\citep{Podesta2012}, namely 
\begin{equation}\label{1}
\delta B_{\perp}(f)={\bf \delta B}(f).\frac{{\bf k}\times {\bf B}_0}{|{\bf k}\times{\bf
B}_0|}\sim {\bf \delta B}(f).\frac{{\bf V}_f\times {\bf B}_0}{|{\bf V}_f\times{\bf
B}_0|}
\end{equation}
Figure~\ref{fig:fig4} shows a comparison between the observed magnetic compressibility (from the data of
Fig.~\ref{fig:fig1}) compared to theoretical ones calculated using the
observed plasma parameters. First, one can see that the theoretical magnetic compressibilities of the fast and slow modes in the
fluid and kinetic models have the same profile being almost constant at the MHD and
sub-ions scales. The KAW mode shows an increasing magnetic compressibility as it
approaches kinetic scales~\citep{Sahraoui2012}. A similar profile has been reported
in solar wind observations~\citep{Podesta2012,Kiyani2013a}. Second,  the measured magnetic compressibility shows a relatively constant and high level ($C_B>1/3$) at the energy containing scales and in the sub-ions range, which indicates the dominance of the parallel component $\delta B_{\parallel}$ (most of the 42 studied intervals showed a similar profile). This
clearly rules out the Alfv\'enic fluctuations as a dominant component of the
turbulence at least at MHD scales ($f <0.05$Hz).

Figure~\ref{fig:fig4} shows that the magnetic compressibility cannot be
used to distinguish between fast and slow modes. To do so, we use instead the
cross-correlation between the magnetic field and the plasma density fluctuations
$C(\delta B_{||},\delta n)$. Indeed, the fast (resp. slow) mode is known to have a
correlation (resp. anti-correlation) between its density and parallel magnetic
component ~\citep{Gary1992}.
\begin{figure}
        \begin{centering}
                \includegraphics[width=1\columnwidth]{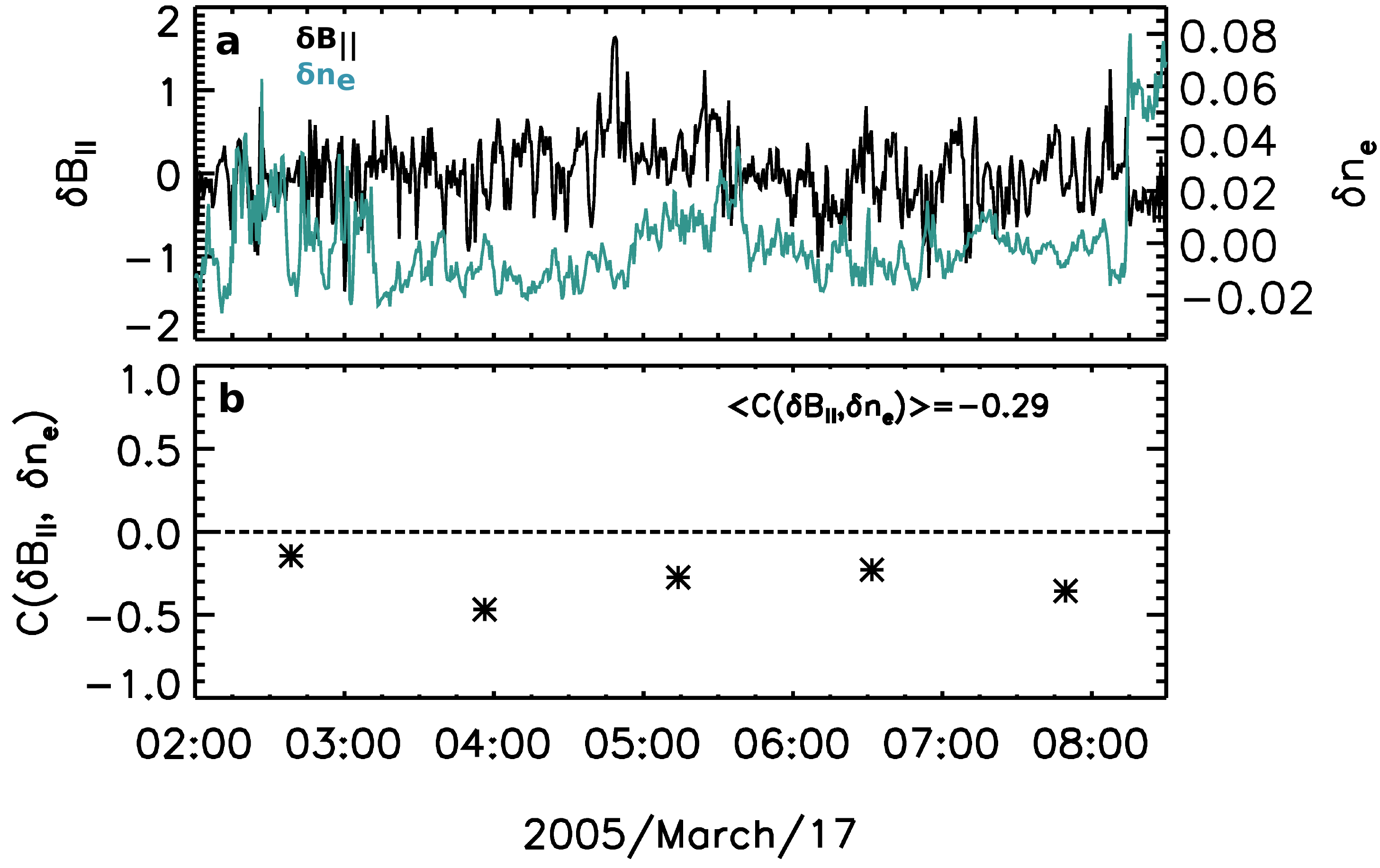}\par
        \end{centering}
        \caption{\label{fig:fig5} (a) The plasma density and the magnetic field
magnitude; (b) The local and averaged cross correlation of the density and
the parallel component of the magnetic fluctuations calculated using
Pearson's method.}
\end{figure}
Figure~\ref{fig:fig5} shows that locally and on average the density and the parallel
component of the magnetic fluctuations are anti-correlated, i.e. $C(\delta
B_{||},\delta n_{e})<0$. This clearly rules out the fast mode fluctuation as the
dominant component of the turbulence. This analysis establishes that the
magnetosonic slow-like mode dominates the turbulent fluctuations analyzed here in
agreement with previous results on the Earth's
magnetosheath~\citep{Kaufmann1970,Song1994,Bavassano2000}, in outer
planets~\citep{Violante1995,Erdos1996} and in the solar wind~\citep{Howes2012,Klein2012}. However, one cannot rule out the possible
presence of the ion mirror mode as previously reported in the terrestrial
magnetosheath~\citep{Sahraoui2006}. The mirror mode, although is of purely kinetic
nature~\citep{Southwood1993}, has indeed very similar properties than the slow mode
which makes it challenging to distinguish between the two modes in spacecraft data. To
check the possibility for the mirror mode to exist in our data requires measuring at
least the ion temperature anisotropy, which is not available onboard the Cassini
spacecraft.
\subsection{Higher order statistics of the magnetic field fluctuations} 
\begin{figure}
        \begin{centering}
                \includegraphics[width=1\columnwidth]{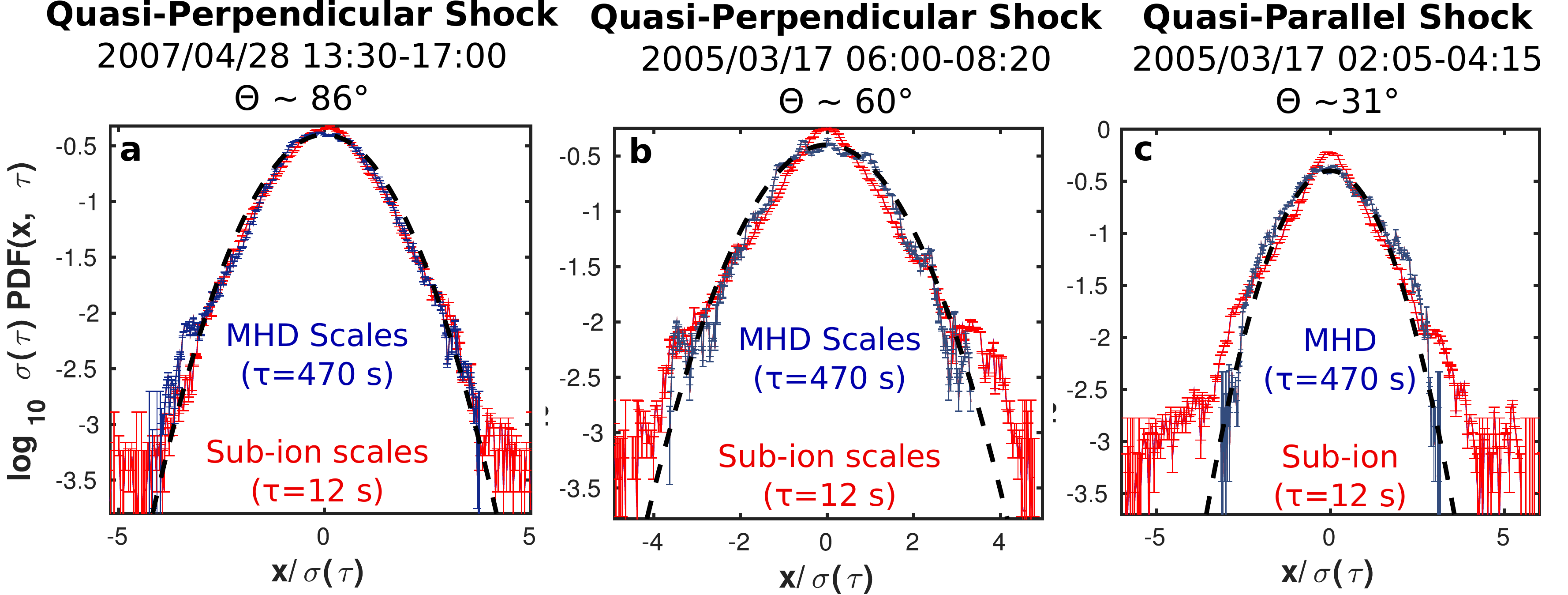} \par
        \end{centering}
        \caption{\label{fig6} PDFs of the magnetic field increments in the energy
containing and sub-ion scales (blue and red respectively) downstream of
quasi-perpendicular (a-b) and quasi-parallel (c) shocks (with Poisonnian
error bars). Normalized histograms with 300 bins each were used to compute
the PDFs. The same values of $\tau$ were used in both cases, $\tau \sim 12$ s
for the kinetic scales and $\tau\sim 470$ s for the MHD scales.  All the
PDFs have been rescaled to have unit standard deviation. A Gaussian
distribution (black dashed curve) is shown for comparison.} 
\end{figure}

\begin{figure}
        \begin{centering}
                \includegraphics[width=1\columnwidth]{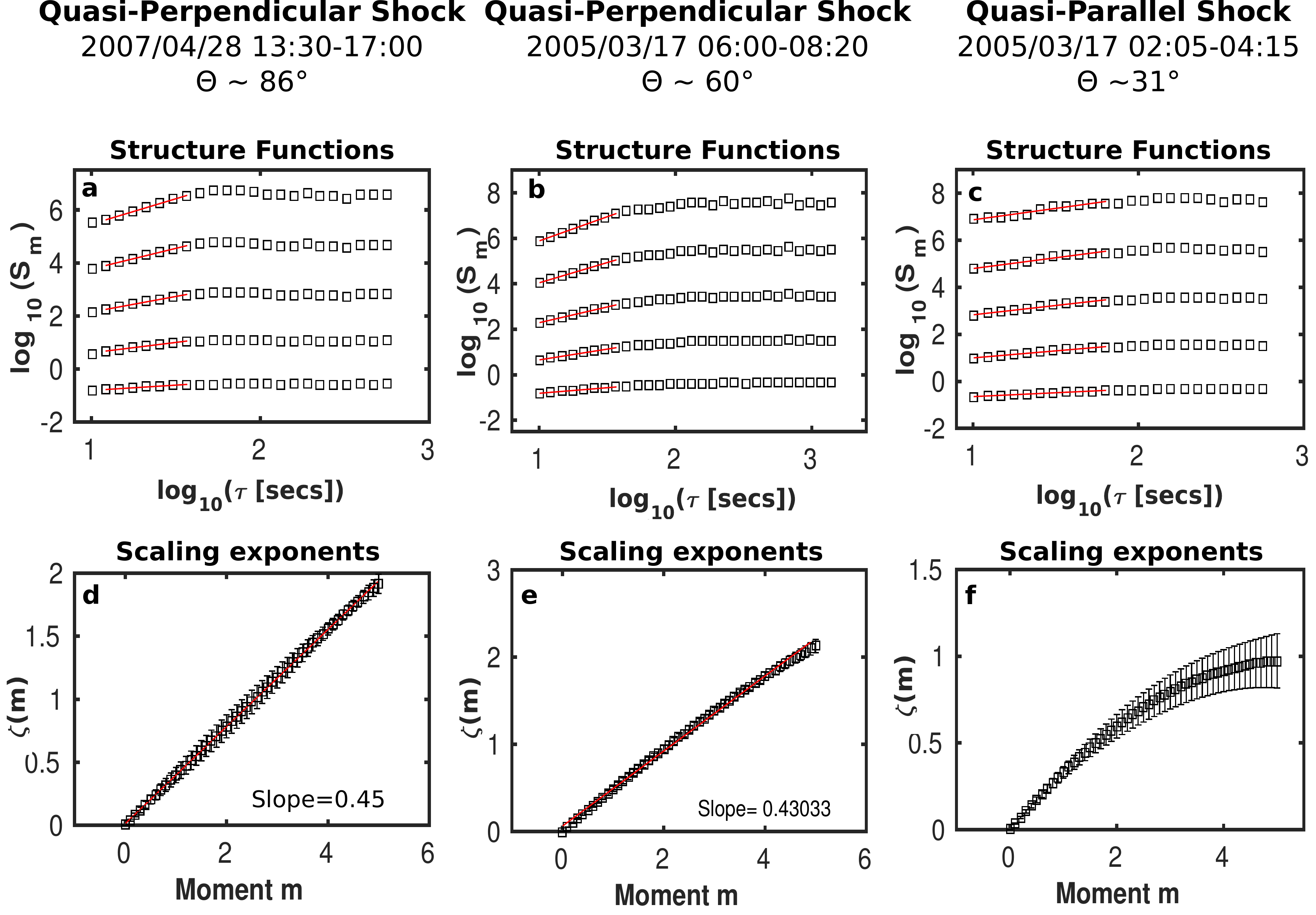} \par
        \end{centering}
        \caption{\label{fig:fig7}Different orders of the structure functions of the
magnetic field increments $\delta B_\tau(t)$ as function of the time lag
$\tau$ downstream of quasi-perpendicular (a-b) and quasi-parallel (c)
shocks.  (d),  (e)  and (f) represent the corresponding scaling exponent
$\zeta(m)$.} 
\end{figure}

To investigate the mono-fractal versus multi-fractal nature of the observed
turbulence we analyzed the Probability Density Function (PDF) of the magnetic field
temporal increments, defined as $\delta B_\tau(t,\tau)=B(t+\tau)-B(t)$ where $\tau$
is the time lag. Intermittency is generally characterized by the presence of bursty
increments which yield heavy tails in the PDF of the field increments at small
scales. In general, it is this deviation from Gaussianity that contains the pertinent
information about the underlying physics.
Figure~\ref{fig6} provides three examples of the corresponding PDFs obtained from the
list of the analyzed events downstream a quasi-parallel ($\Theta=31^\circ$) and quasi-perpendicular shocks ($\Theta=86^\circ$ and $\Theta=60^\circ$, respectively) for two values of $\tau$, one from the MHD range ($\tau \sim 975$ s) and one from 
the sub-ion range ($\tau \sim 25$ s). 
Figures~\ref{fig6}-(a-b) show that behind the quasi-perpendicular shock, the PDFs are
found quasi-Gaussian for both values of $\tau$ (i.e., at MHD and kinetic scales)
indicating the quasi-randomness of the fluctuations in the ``energy containing sacles" and in the
sub-ion range. On the contrary, behind  the quasi-parallel shock
(Figure~\ref{fig6}-c), the PDFs are non-Gaussian for $\tau \sim 25$ s (red PDF)
showing the intermittent nature of turbulence at the kinetic scales as it was
observed in the terrestrial magnetosheath~\citep{Sundkvist2007,Yordanova2008}. In
the energy containing scales the PDF is quasi-Gaussian as in the case of
quasi-perpendicular shock. These results agree with recent findings using global
hybrid simulations~\citep{Karimabadi2014}.
We next calculate higher order statistics given by the structure functions (SFs) of
the magnetic field increments defined in Eq.~\ref{2}.
\begin{equation}
S_{m}(\tau)=\intop_{-\infty}^{\infty}|\delta B_{\tau}(t)|^{m}P(\delta B)dt=<|\delta
B_{\tau}(t)|^{m}>\, \label{2}
\end{equation}
When increasing the order $m$, the SFs become progressively sensitive to rare and
bursty events. Assuming a power-law dependence of the SFs as function of $\tau$,
i.e., $S_{m}(\tau)\sim\tau^{-\zeta(m)}$, a linear (resp. nonlinear) dependence of
$\zeta(m)$ on the order $m$ indicates a mono-fracatal (resp. a multi-fractal)
behaviour of the turbulence. The scaling exponent in Figure~\ref{fig:fig7}-a shows a
clear linear dependence of $\zeta$ on $m$ at the sub-ion scales ($\equiv$ small
values of $\tau)$ behind a quasi-perpendicular shock supporting thus the monofractal
character of the turbulent fluctuations.
 However, downstream a quasi-parallel shock (Figure~\ref{fig:fig7}-c), the scaling exponent
is a convex function of ${m}$, confirming the multifractal nature of turbulence at
the kinetic scales. We recall that in the solar wind it has been shown that the
sub-ion scales were monofractal while the MHD scales were
multi-fractal~\citep{Kiyani2009a}.
\section{Discussion and conclusions} \label{sec:Conclusions} 
From the benefit of analyzing sufficiently long and relatively stationary time
series measured by the Cassini spacecraft in the magnetosheath of Saturn, we were
able to probe into more than four decades of scales spanning from the MHD down to
the sub-ion scales. We present the following plausible, albeit speculative, scenario to explain the different
observations: the interaction of the solar wind with the bow shock may lead to the
destruction of all the pre-existing  correlations between the turbulent fluctuations
in the solar wind. This results in suppressing the Kolmogorov inertial range and
generating locally random-like fluctuations that have a scaling $\sim f^{-1}$ over a
broad range of scales. Those scales would play the same role as the energy containing scales
in solar wind turbulence. The absence of the inertial range scale in our
observations can be explained by the fact that the newly generated fluctuations
behind the shock do not have ``enough time'' to interact sufficiently with each
other to reach a fully developed turbulence state, hence the direct transition from
the ``energy containing" range that has $\sim f^{-1}$ scaling into the sub-ion range
with a scaling $\sim f^{-2.6}$. In this scenario, turbulence may reach a fully
developed state and the Kolmogorov $5/3$ spectrum may be observed but only far away
from the shock (e.g., toward the flanks).
However, a fundamental question remains open: How the power-law spectra observed in
the sub-ion range are created in the absence of an inertial range? We note that
existing theoretical models of kinetic turbulence in the solar wind suggest that
turbulence at sub-ion scales may result as a consequence of a decoupling between the
dynamics of Alfv\'enic fluctuations and the rest of the MHD fluctuations that would
carry the cascade into the kinetic scales~\citep{Schekochihin2009}. Hence, if the
MHD scales are not dominated by Alfv\'enic fluctuations as observed here, it is not
clear as to how turbulence is generated at kinetic scales. Is it generated by local
plasma instabilities occurring near the ion scale as proposed
in~\cite{Sahraoui2006}? In this case, would the $f^{-1}$ spectrum observed at larger
scales result by an inverse cascade as observed in hydrodynamic
turbulence~\citep{Paret1997,Chertkov2007}? Another observation reported here that
requires further investigations is the nature of the turbulence observed behind the
quasi-perpendicular shock. At sub-ion scales, turbulence was found to have a
mono-fractal behaviour as in the solar wind~\citep{Kiyani2009b}. However, the tails
of the PDFs are clearly less pronounced in our observations and the PDF of
Figure~\ref{fig6}-a looked close to a Gaussian rather than a heavy tailed PDF.  This
result recalls recent numerical observations of weak turbulence in
Electron-MHD~\citep{Meyrand2015}. This similarity and the questions raised above
require further investigation in the future.
\begin{acknowledgments}
Authors thank the NASA's Planetary Data System (PDS) for the CASSINI-CAPS data set
availability,  (Waite, J.H., Furman, J.D., CASSINI ORBITER SAT/SW CAPS DERIVED ION
MOMENTS V1.0,CO-S/SW-CAPS-5-DDR-ION-MOMENTS-V1.0, NASA Planetary Data System, 2013).
Authors are also indebted to the program ``Soleil H\'eliosph\`ere et
Magn\'etosph\`eres" of CNES, the french space administration for the financial
support on CASSINI. Data analysis were done with the AMDA science analysis system
provided by the Centre de Donn\'ees de la Physique des Plasmas (IRAP, Universit\'e
Paul Sabatier, Toulouse) supported by CNRS and CNES. LH thanks G. Belmont for many
fruitful discussions.
\end{acknowledgments}

\bibliographystyle{apj}

\end{document}